\documentclass[aps, superscript address,float fix,one column,reprint,footinbib]{revtex4-1} 

\usepackage{amsmath,amssymb,bm,epsfig}
\usepackage{color}
\usepackage{natbib}
\usepackage{hyperref} 
\usepackage{ulem}
\usepackage{graphicx}
\usepackage{wasysym}
\RequirePackage{lineno}

\usepackage[utf8]{inputenc}

\usepackage{xcolor,colortbl}
\usepackage{pifont}
\usepackage{footnote}

\def \beq{\begin{equation}}
\def \eeq{\end{equation}}
\def \beqa{\begin{eqnarray}}
\def \eeqa{\end{eqnarray}}
\def \la{\langle}
\def \ra{\rangle}

\newcommand{\xt}{{\mathbf{x}_\perp}}

\begin{document}

\title{Mass ordering of spectra from fragmentation of saturated gluon states in high multiplicity proton-proton collisions}

\author{Bj\"orn Schenke}
\affiliation{Physics Department, Brookhaven National Laboratory, Upton, NY 11973, USA}

\author{S\"oren Schlichting}
\affiliation{Physics Department, Brookhaven National Laboratory, Upton, NY 11973, USA}

\author{Prithwish Tribedy}
\affiliation{Physics Department, Brookhaven National Laboratory, Upton, NY 11973, USA}

\author{Raju Venugopalan}
\affiliation{Physics Department, Brookhaven National Laboratory, Upton, NY 11973, USA}
\affiliation{Institut f\"{u}r Theoretische Physik, Universit\"{a}t Heidelberg, Philosophenweg 16, 69120 Heidelberg, Germany
}

\begin{abstract}
The mass ordering of mean transverse momentum $\left<p_T\right>$ and of the Fourier harmonic coefficient $v_2 (p_T)$ of azimuthally anisotropic particle distributions in high energy hadron collisions is often interpreted as evidence for the hydrodynamic flow of the matter produced. We investigate an alternative initial state interpretation of this pattern in high multiplicity proton-proton collisions at the LHC. The QCD Yang-Mills equations describing the dynamics of saturated gluons are solved numerically with initial conditions obtained from the Color Glass Condensate based IP-Glasma model. The gluons are subsequently fragmented into various hadron species employing the well established Lund string fragmentation algorithm of the PYTHIA event generator.  We find that this initial state approach reproduces characteristic features of bulk spectra, in particular the particle mass dependence of $\left<p_T\right>$ and $v_2 (p_T)$. 
\end{abstract}
\maketitle

It is now well established that the QCD matter formed in collisions of heavy nuclei (A+A) behaves like a strongly interacting fluid that exhibits collective features described by the equations of relativistic viscous hydrodynamics \cite{Gale:2013da}. 
A striking recent finding is that some observables measured in high multiplicity events of much smaller collision systems like $p+p, p+A, d+A$ and $^3\!He+A$ resemble features of $A+A$ collisions that are attributed to collective flow of the produced matter~\cite{Khachatryan:2010gv, CMS:2012qk, Abelev:2012ola, Aad:2012gla, Adare:2014keg, Adamczyk:2015xjc, Adare:2015ctn, Khachatryan:2015lva, Aad:2015gqa}. However these small systems also contain puzzling features that are not easily reconciled with collectivity; an example is the pronounced back-to-back azimuthal angle correlation of di-hadrons in small systems~\cite{Khachatryan:2016txc}, which is significantly quenched for the larger systems \cite{Adler:2002tq}. 

Since hydrodynamics relies on a separation of the macroscopic and microscopic scales, it is very interesting to explore what are the smallest size systems that can be efficiently described as hydrodynamic fluids. 
{Collective effects in p+p collisions have been discussed for some time--see for instance, Ref.~\cite{Levai:1991be}.} While hydrodynamics and kinetic theory describe some of the trends in the data~\cite{Bozek:2013uha,Bzdak:2013zma,Qin:2013bha, Werner:2013ipa, Schenke:2014zha,Bzdak:2014dia, Kalaydzhyan:2015xba}, this apparent efficacy of hydrodynamics in small systems outstretches simple estimates of its applicability~\cite{Niemi:2014wta}. 
Conversely, if final state interactions are weak, the correlations attributed to hydrodynamic behavior in small systems may provide insight into many-body correlations in the initial state and their non-equilibrium dynamical evolution. 

Initial state descriptions based on the Color Glass Condensate (CGC) effective theory~\cite{Gelis:2010nm}  of the  initial non-equilibrium Glasma~\cite{Gelis:2006dv,Lappi:2010ek} of highly occupied gluon states, provide qualitative~\cite{Dumitru:2010iy,Kovner:2010xk, Dusling:2012iga,Kovchegov:2012nd,Dumitru:2014dra,Dumitru:2014yza,Gyulassy:2014cfa, Lappi:2015vha} and semi-quantitative~\cite{Dusling:2012cg, Dusling:2012wy, Dusling:2013qoz, Schenke:2015aqa, Lappi:2015vta, Dusling:2015rja} descriptions of several features of  small systems. { While the CGC may provide an appropriate description of rare, high multiplicity gluon states,  a shortcoming
of current computations is that they either directly compare gluon distributions to data or employ fragmentation functions that are not reliable at the  
low $p_T$ where 
collective dynamics should be dominant~\cite{Esposito:2015yva}.} These computations were thus unable to address the particle species dependence of the average transverse momentum $\la p_T\ra$ as a function of multiplicity and that of the Fourier harmonic $v_2(p_T)$ in high multiplicity events - the observed mass-splitting patterns of both these quantities were previously adduced as strong evidence of hydrodynamic flow~\cite{Bozek:2013ska}. 

In this letter, we will introduce a CGC+Lund model that provides a mechanism to fragment gluons emerging from the Glasma into various hadron species and apply the model to address the question whether the aforementioned mass-splitting patterns can be reproduced in an initial state framework.  We note that there have been previous merging of CGC based models in $k_T$ factorization frameworks~\cite{Deng:2014vda, Albacete:2016tjq} to string fragmentation. However, our approach is the first fully dynamical one, combining the IP-Glasma model of event-by-event Yang-Mills evolution of primordial color charge fluctuations~\cite{Schenke:2012wb,Schenke:2012hg}, with the state-of-the-art Lund string fragmentation algorithm~\cite{Andersson:1983ia, Sjostrand:1984ic} of the PYTHIA event generator~\cite{Sjostrand:2014zea}. 
This novel CGC+Lund Monte-Carlo event generator will enable us in the future to address a wide range of phenomenological questions concerning small systems. 

\textit{Implementation:}  
Our framework is based on the IP-Glasma model \cite{Schenke:2012wb,Schenke:2012hg}, which provides a successful description of multi-particle production in high energy hadronic collisions. In the IP-Glasma model,  the spatial proton shape $T_{p}(\xt)$ is a Gaussian distribution, with its width constrained by a comparison of exclusive deeply inelastic scattering (DIS) data from HERA to the IPSat model~\cite{Rezaeian:2012ji}. The overall normalization of the color charge density distribution is proportional to the saturation scale $\langle Q_{S}(\xt,x)\rangle$ inside the proton, which is determined self-consistently from the relation $x=0.5\, \langle Q_S(\xt,x)\rangle/\sqrt{s}$~\cite{Schenke:2013dpa} at a given collision energy $\sqrt{s}$. Here we also incorporate intrinsic fluctuations of the proton saturation scale according to the distribution \cite{McLerran:2015qxa}
\beq
P(\ln(Q_S^2/\langle Q_S^2 \rangle)) =\frac{1}{\sqrt{2\pi}\sigma}\exp\left(  -\frac{\ln^{2}(Q_{S}^2/\langle Q_{S}^{2} \rangle)}{2\sigma^{2}}\right),%
\label{eq_qsdist}
\eeq
where 
$\sigma=0.5$ 
is constrained by  
independent analyses using experimental data of inclusive charged particle multiplicity and rapidity distributions in p+p collisions \cite{McLerran:2015qxa, Bzdak:2015eii} and 
by HERA DIS data~\cite{Mantysaari:2016ykx}. Events with $Q_S^2/\langle Q_S^2\rangle > 1$ correspond to rare Fock space configurations inside the two colliding protons.
Additional geometric fluctuations of the proton substructure \cite{Schlichting:2014ipa,Mantysaari:2016ykx} 
have only a small effect on the intrinsic azimuthal correlations we will study here \cite{Schenke:2015aqa}.

In each event, the collision geometry is determined by sampling impact parameters according to an eikonal model for p+p collisions~\cite{d'Enterria:2010hd}. Subsequently, color charges inside the two protons $\rho^a(\xt)$ are sampled from a Gaussian distribution according to the MV model~\cite{McLerran:1993ni,McLerran:1993ka} 
\begin{equation}
\langle \rho^a(\xt)\rho^b({\rm y}_\perp)\rangle = g^2\mu^2(\xt,x) \delta^{ab} \delta^{(2)}(\xt-{\rm y}_\perp) \, ,
\label{eq_mv}
\end{equation}
where $g^2\mu(\xt,x)$ = $Q_S(\xt,x)/\xi$. The parameter $\xi$ is a non-perturbative constant, which 
has been constrained by global analysis of multiplicity distributions in different collision systems~\cite{McLerran:2015qxa, Schenke:2013dpa} to be 
$0.45\le \xi \le 0.75$. 

Numerical solutions of the classical Yang-Mills equations 
determine the color fields for each configuration of color charges~ \cite{Schenke:2012wb,Schenke:2012hg}. In our study, the gluon fields produced after the collision are evolved 
up to time $\tau\sim 1/Q_S$.  The multiplicity density $dN_g/dyd^{2}k_{T}$ of produced gluons is then determined in Coulomb gauge \cite{Krasnitz:2001qu, Lappi:2003bi}. On an event by event basis, the fluctuations of the color fields inside each colliding proton induce azimuthal correlations of the gluons produced on time scales $\sim 1/Q_S$ \cite{Schenke:2015aqa}. Because $dN_g/dy  d^2k_T \sim 1/g^2$, the effect of a running coupling can be introduced by multiplying 
the gluon multiplicity with an effective factor of $g^2/(4\pi \alpha_s(\tilde{\mu}))$, where the scale for running $\tilde{\mu}$ is chosen to be 
the gluon transverse momentum $k_T$. Unless otherwise noted the results presented here employ running coupling. The absolute normalization of the gluon density $dN_g/dyd^{2}k_{T}$ is sensitive to the choice of the lattice parameters in our study and the running coupling scheme~\footnote{We will use the lattice parameters $N=400$, $L=8$ fm, $\tau_{\rm evol}=$ 0.4 fm, $g^2\mu =0.45$ and an infrared regulator of $m=0.1$ GeV.   For these numerical details, and the sensitivity of 
results to variations of these, we refer the reader to the extensive study in Ref.\,\cite{Schenke:2013dpa}.}.
To circumvent this ambiguity, we will express our results in terms of the scaled multiplicity $dN_g/dy/ \la dN_g/dy\ra$.

\begin{figure}[t]
\includegraphics[width=0.45\textwidth]{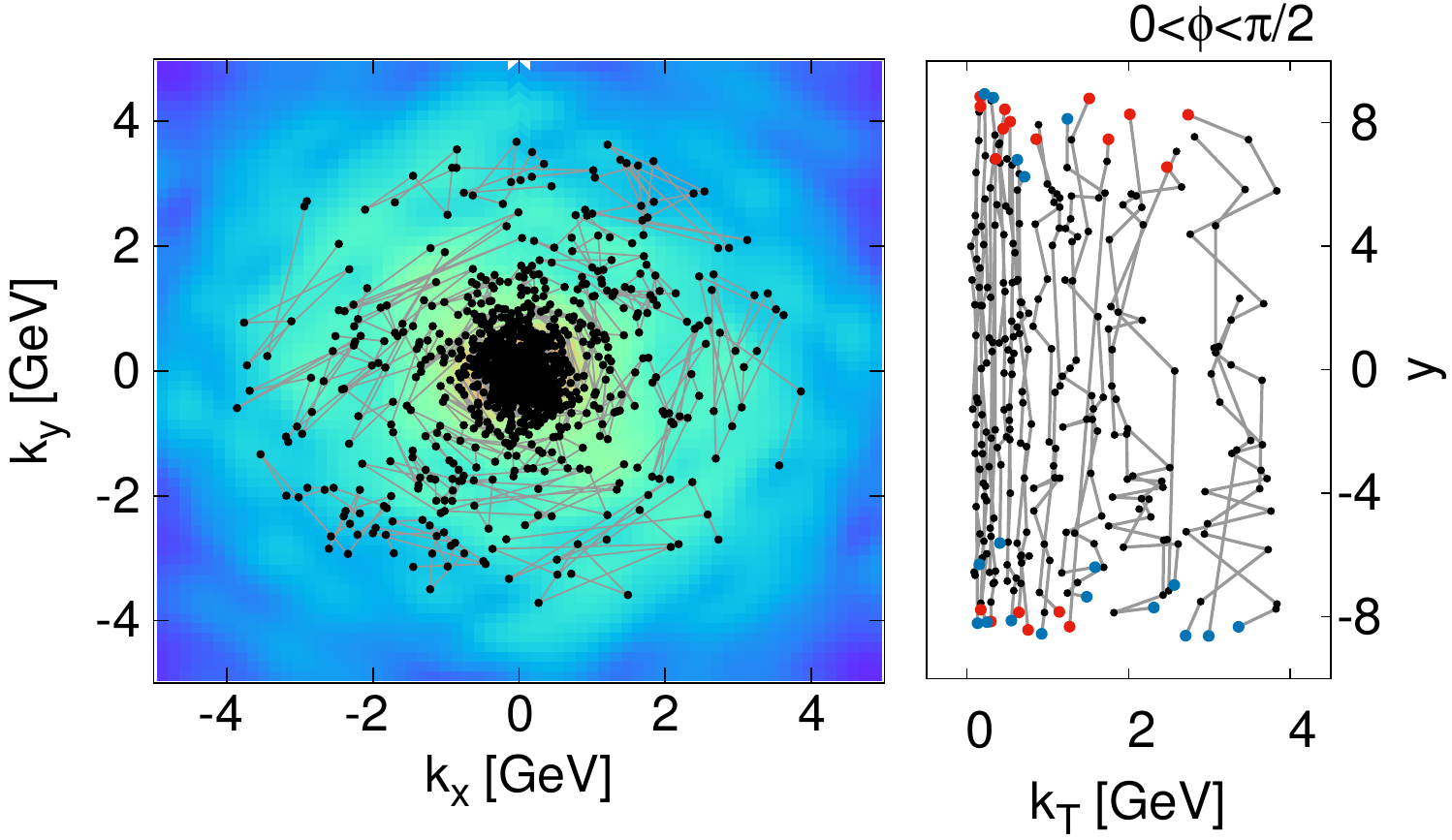} 
\caption{Left: Initial gluon distribution for a single IP-Glasma event; a single configuration of the sampled gluons in momentum space is shown by black points. Right: Strings extended in rapidity and clustered in transverse momentum. Red and blue points represent the momentum of the quarks and anti-quarks attached at the string ends. Strings connecting the sampled gluons are shown by grey lines.} 
\label{fig_gludist} 
\end{figure}

In order to pass the IP-Glasma events to the string fragmentation algorithm implemented in PYTHIA, we sample $N_g$ gluons, where $N_g$ is determined by integrating $dN_g/dyd^{2}k_{T}$ over a range of rapidity $\Delta y_{\rm max}$ and transverse momentum $k_{\rm _{T}, max}$.
The rapidity of these gluons is sampled from a uniform distribution over the range $-y_{\rm max} < y <-y_{\rm max}$, and the transverse momentum is sampled from $dN_g/d^{2}k_{T}$. 
The maximum value of rapidity $y_{\rm max}$ 
is equal to the beam rapidity of the colliding protons, $y_{\rm max} =\log(\sqrt{s}/m_{p})=8.9$ at 7 TeV, with $m_{p}$ the proton mass. We choose the maximum transverse momentum of the sampled gluons to be $k_{\rm T, max}=10$ GeV.  We then feed the momentum and color structure of the sampled gluons into PYTHIA's particle list \cite{Sjostrand:1985xi} by constructing strings. Each sampled gluon in the event is assigned a color index;  a fixed number of these are grouped together into 
a single string inspired by the Glasma flux tube picture~\cite{Dumitru:2008wn}. We use a fixed value of $18$ for the number of gluons in each string $N_{\rm gs}$, which corresponds to the average value of $N_g/\la Q_S^2 S_\perp \ra$, where $\la Q_S^2 S_\perp \ra$ is the number of flux tubes and $S_\perp$ denotes the transverse overlap area. 

We group gluons close in transverse momentum space into strings stretching mainly in the rapidity direction and add a quark and an anti-quark at 
string ends to guarantee 
color neutrality
~\footnote{The quark and the anti-quark are chosen to be massless and have 3-momenta equal to the gluons attached to them, ensuring they do not contribute to particle production. At any rate, by construction, the quark and the anti-quark have large rapidities and should hence not affect results close to mid-rapidity.}. In Fig.\ref{fig_gludist} (left) we show the momentum space distribution of the initial gluon density obtained from the IP-Glasma model at time $\tau=0.4$ fm together with the positions of the sampled gluons; the configuration of the PYTHIA strings is shown in Fig.\ref{fig_gludist} (right). 

In this work, we will use the ``hadron-stand-alone-mode" of PYTHIA, which employs the 
Lund symmetric fragmentation function 
\beq
f(z, m_{T}) = \frac{1}{z} (1-z)^{a} \exp \left(-\frac{b\, m{_T}^2}{z} \right)\;.
\eeq
Here $m_{T}$ and $z$ denote the transverse mass and the light cone momentum fraction of the fragmenting hadron, and the default parameters $a=0.68$ and $b=0.98$ are constrained by a global data analysis~\cite{Sjostrand:2014zea}. Further, the transverse momenta of hadrons during the fragmentation are smeared according to a Gaussian distribution with the width $\sigma_{p_T}\!=\!0.33$ GeV. 
We do not modify 
the default parameters in PYTHIA for our study. Variations of our results with respect to these parameters are discussed in the supplementary material. 
In order to acquire sufficient statistics,  we generate fifty sampled gluon (and string) configurations from every IP-Glasma event, and hadronize each gluon configuration 100 times.

Unless otherwise noted, our results include the color reconnection procedure of PYTHIA, which can be enforced after initializing the string configurations. 
\begin{figure}[t]
\includegraphics[width=0.45\textwidth]{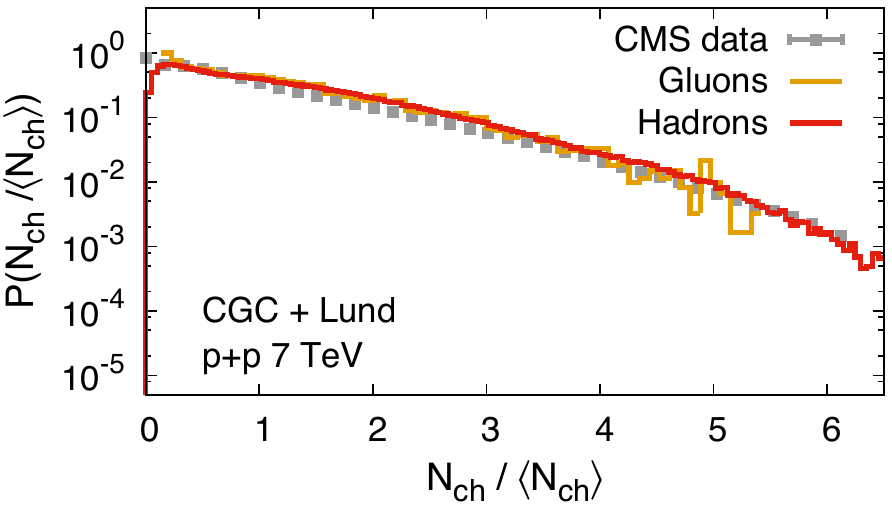}
\caption{Probability distribution of scaled charged hadron multiplicity measured over $|\eta|<0.5$ in p+p collisions at 7 TeV. The data points are from Ref.~\cite{Khachatryan:2010nk}.} 
\label{fig_multdist} 
\end{figure}

{\it Results:}
We will restrict ourselves here to bulk observables in p+p collisions at  $\sqrt{s}=7\,{\rm TeV}$. In Fig.\,\ref{fig_multdist} we compare the probability distribution of the scaled gluon multiplicity and the inclusive hadron multiplicity to experimental data on inelastic non-single diffractive events from the CMS collaboration \cite{Khachatryan:2010nk}. We note that the IP-Glasma model naturally produces multiplicity distributions of gluons that are a convolution of multiple negative binomial distributions ~\cite{Gelis:2009wh,Schenke:2012wb,Schenke:2012hg}. 
In computing the multiplicity distribution, we included all events in which the rapidity density of gluons $dN_{g}/dy \ge 1$~\footnote{Including all events regardless of $dN_{g}/dy$ reduces the mean multiplicity by 16$\%$.}. The multiplicity of charged hadrons $dN_{\rm ch}/dy$ is about $50-75\%$ larger than $dN_{g}/dy$ depending on the coupling used. Fragmentation however does not significantly change the shape of the distribution of the scaled multiplicity. Within the available statistics we find very good agreement with the data up to six times the mean multiplicity.

\begin{figure}[t]
\includegraphics[width=0.44\textwidth]{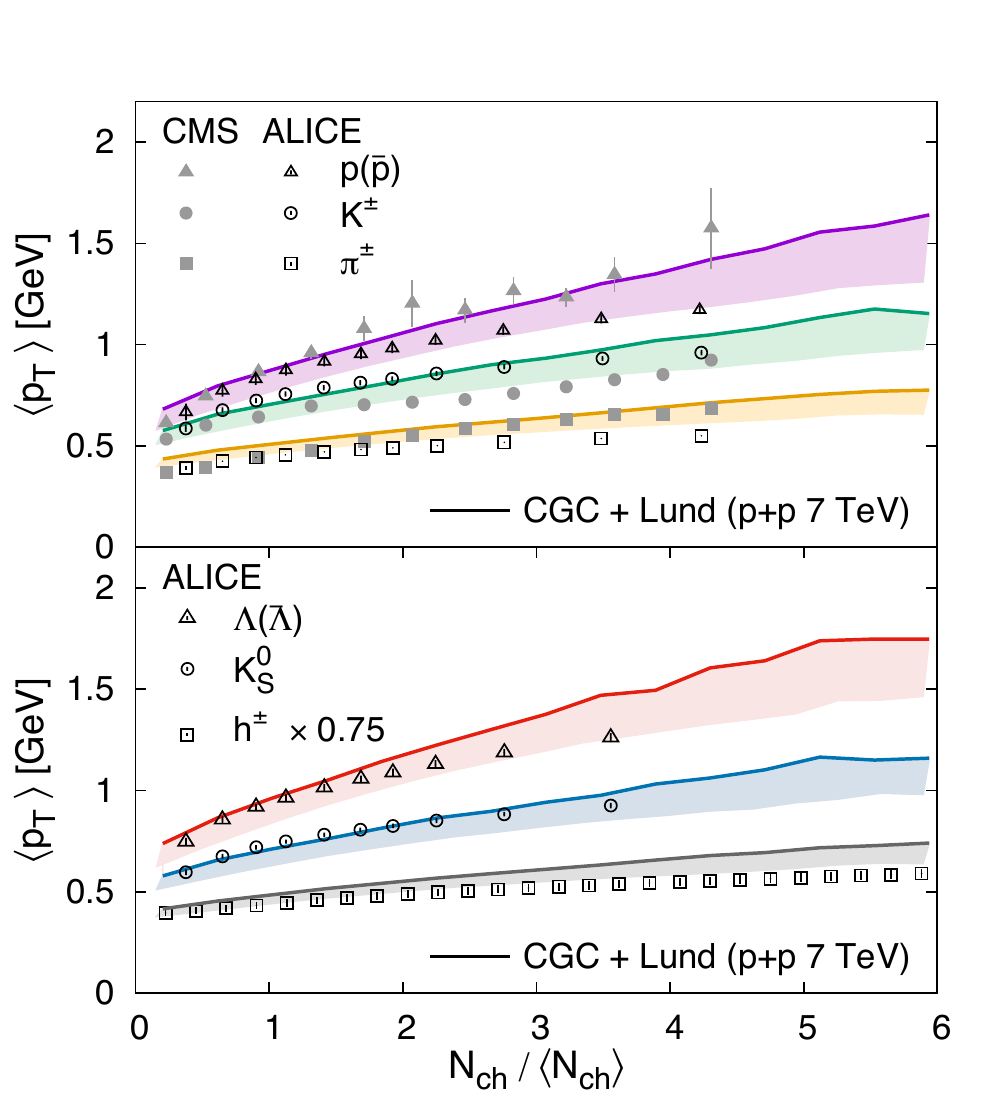}
\caption{Mass ordering of $\la p_T \ra$ plotted against scaled charged hadron multiplicity $N_{\rm ch}/\la N_{\rm ch}\ra$. Data points for identified particles from the ALICE~\cite{Bianchi:2016szl} and CMS\,~\cite{Chatrchyan:2012qb} Collaborations are in the range $|y|<0.5$ and $|y|<1$, respectively. The values corresponding to $\langle N_{\rm ch} \rangle$ are obtained from Ref.~\cite{Adam:2015gka} and Ref.~\cite{Khachatryan:2010nk} for ALICE and CMS data correspondingly. The $\la p_T \ra$ values for charged hadrons are obtained from Ref.~\cite{Abelev:2013bla}.}  
\label{pid_avgpt}
\end{figure}

We now present results for the average transverse momentum $\la p_T \ra$ for charged hadrons over the experimentally used range of transverse momentum $0.15\,{\rm GeV} < p_{T} < 10.0$ GeV and $|\eta|<0.3$, and for identified hadrons $\pi^{\pm}, K^{\pm}, p(\bar{p}), K_{S}^0$ and $\Lambda/(\bar{\Lambda})$ for a rapidity range of $|y|<0.5$, with no cuts on transverse momentum. 
We compare our calculation to the  preliminary and published measurements from the ALICE \cite{Abelev:2013bla, Bianchi:2016szl} and CMS collaborations \cite{Chatrchyan:2012qb}. To perform a consistent comparison between data and our computation, we show the variation of $\left< p_T \right>$ with the scaled charged hadron multiplicity $N_{\rm ch}/\left< N_{\rm ch}\right>$ in Fig.\,\ref{pid_avgpt}~\footnote{The value of the mean multiplicity in the ALICE data over $|\eta|<0.5$ is $\left<N_{\rm ch}\right>$=5.98~\cite{Adam:2015gka}, and for the CMS data over $|\eta|<2.4$ $\left<N_{\rm ch}\right>$=30 ~\cite{Khachatryan:2010nk}. For ALICE we have $\left<N_{\rm ch}\right>$=4.42 in the range $|\eta|<0.3$\cite{Abelev:2013bla}.}.

Our results for the multiplicity dependence of $\la p_T \ra$ in the 
running coupling case are shown by solid lines. The bands shown include the variation due to using fixed coupling, which decreases $\la p_T \ra$ by about $10-15\%$ 
and the effect of turning off color reconnections in PYTHIA fragmentation which decreases $\la p_T \ra$ by about $5-10\%$~\footnote{By way of contrast, in standard PYTHIA, the color reconnection at parton level increases the $\left< p_T \right>$ by 40-60$\%$~\cite{Abelev:2013bla}.}. We see a strong increase of $\la p_T \ra$ with increasing multiplicity, consistent with the data. More interestingly, we find that our framework naturally reproduces the mass ordering for different species: $\la p_T \ra_{\!_p} \!\!>\!\! \la p_T \ra_{\!_K} \!\!>\!\! \la p_T \ra_{\!_\pi} $ and $\la p_T \ra_{\!_\Lambda} \!\!>\!\!\la p_T \ra_{\!_{K_S^0}} \!\!>\!\! \la p_T \ra_{\!_h} $ over the entire range of multiplicity considered. 

The strong multiplicity dependence of $\la p_T \ra$ and the mass ordering {was demonstrated to arise in the fragmentation of mini-jets in HIJING calculations~\cite{Wang:1991vx}. Such effects are also obtained in PYTHIA calculations within the color-reconnection scheme~\cite{Sjostrand:2013cya,Abelev:2013bla,Ortiz:2013yxa,Bierlich:2015rha}}. In PYTHIA, high multiplicity events are associated with a large number of independent parton showers.
These hadrons fragmenting from independent showers will have $\la p_T \ra$ to be independent of the number of showers and therefore independent of $\la N_{\rm ch}\ra$. The inclusion of color-reconnections modifies this by generating correlations between partons from different showers; this leads to collective hadronization of strings and the  strong correlation observed between $\la p_T \ra$ and $\la N_{\rm ch}\ra$. 

Both parton showering
and multi-parton interactions 
are included in the CGC framework, and all the parton ladders in rapidity, localized within a transverse area $\sim 1/Q_S^2$ are correlated. Specifically, with regard to the correlation between multiplicity $N_{\rm g} \sim Q_S^2 S_{\perp}$  and mean transverse momentum of gluons $\la p_T \ra\sim Q_S$, one finds $\la p_T \ra \sim \sqrt{N_{\rm g}/S_\perp}$ showing that the correlation between $\la p_T \ra$  and $N_{\rm g}$ is already present at the gluon level. {Conversely, the mass ordering of the $\la p_T \ra$ of different species can be attributed to the fragmentation scheme implemented in the hadron-stand-alone mode of PYTHIA. Color-reconnection only has a small effect, because gluons are not associated with separate showers and are already assigned to strings depending on their momenta.} 

The hardening of the transverse momentum distribution and mass ordering of $\la p_T \ra$ are often attributed to strong final state rescattering and collective expansion of a system, and even adduced  as such~\cite{Werner:2013ipa, Kalaydzhyan:2015xba} for these patterns in small collision systems{ ~\cite{Abelev:2013haa, Levai:1991be}}. However, what must give pause to such interpretations is the presence of mass-splitting even for the lowest multiplicity bin, well below $N_{\rm ch}/\la N_{\rm ch}\ra<1$. Our results in Fig.~\ref{pid_avgpt} provide an alternative initial state interpretation for this pattern, at least for the range of $N_{\rm ch}/\la N_{\rm ch}\ra$ considered. 

\begin{figure}[t]
\includegraphics[width=0.45\textwidth]{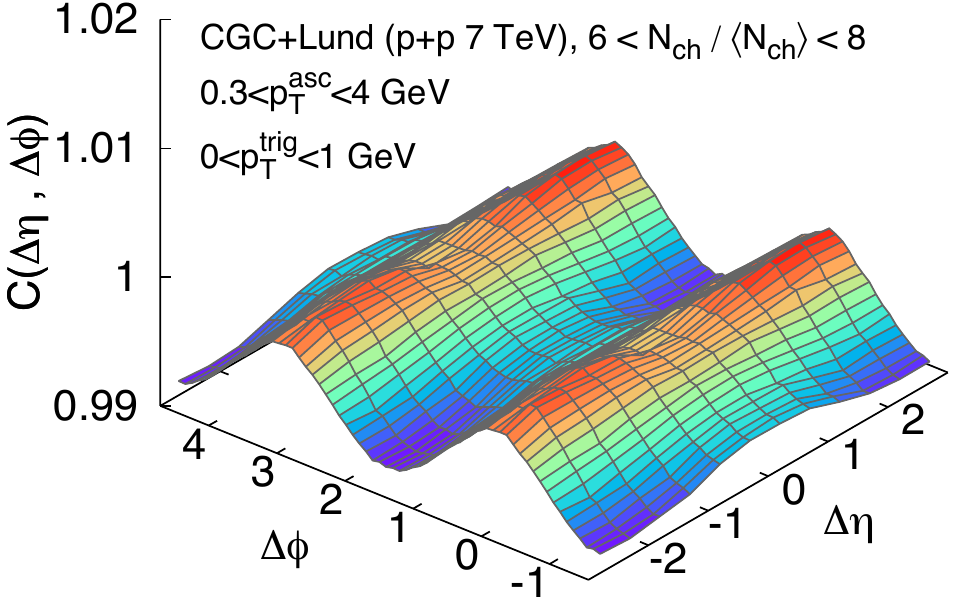}
\caption{Di-hadron correlation function using inclusive charged hadrons as both trigger and associated particles. }
\label{fig_ridge}
\end{figure}

We now extend our considerations to the anisotropy coefficient $v_{2}$ extracted from two particle ``ridge" correlations that are long range in rapidity. The particle species dependence of $v_{2}$ in high multiplicity $p+p$ collisions has been measured recently by the CMS collaboration~\cite{CMS:2015kua, CMS:2016yew}. As a first step, we estimate the two particle correlation function using identified hadrons as trigger particles and inclusive hadrons as associated particles. We hew as closely as possible to the experimental procedure. 
Our analysis is however very computationally intensive and it is challenging to acquire statistics commensurate to those of the experimental data. We compensate by choosing a wider range of pseudo-rapidity $-4<\eta<4$ relative to CMS ($-2.4<\eta<2.4$) but keep the other kinematic cuts identical to the experiment. This includes the range of transverse momentum for associated particles ($0.3 <p_T<3$ GeV) and the range of the relative difference in pseudo-rapidity ($2<|\Delta\eta|<4.8$). The CMS analysis of azimuthal correlations for $\sqrt{s}=7$ TeV was performed  for events with multiplicities ten times higher than the mean multiplicity $\la N_{\rm ch} \ra$. Due to the challenge of acquiring sufficient statistics, we will restrict out study to events with  multiplicities up to eight times the mean multiplicity $\la N_{\rm ch} \ra$.

\begin{figure}[t]
\includegraphics[width=0.45\textwidth]{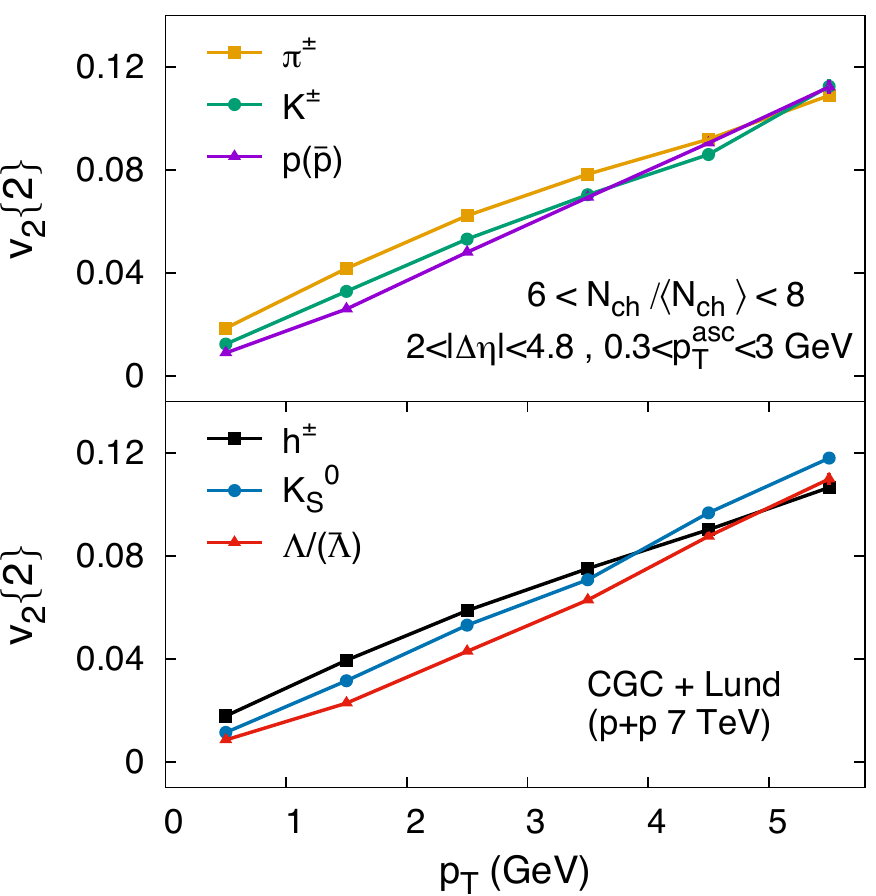}
\caption{Mass ordering of $v_{2}\{2\} (p_{T})$ extracted from the azimuthal dependence of two particle ridge correlations in the CGC+Lund framework. }
\label{fig_v2mass}
\end{figure}

{ The two-particle correlation function $C(\Delta\eta, \Delta\phi)$ \cite{Aad:2012gla} obtained for inclusive charged hadrons in our framework is shown in Fig.\,\ref{fig_ridge}.  A double-ridge structure can clearly be observed. The particle pair distribution function obtained from $C(\Delta\eta, \Delta\phi)$} can be decomposed in a Fourier series of their relative azimuthal angle $\Delta \phi$ as 
\beq
\frac{2\pi}{N_{\rm trig}^{\rm PID} N_{\rm assoc}^{h^\pm}} \frac{dN^{\rm pair}}{d\Delta\phi} = 1 + \sum \limits_{n} 2 V_{n\Delta} \cos (n \Delta \phi), 
\label{eq_v2d}
\eeq
where $N_{\rm trig}^{\rm PID}$ is the number of identified trigger particles (such as $\pi^\pm, K^\pm, p/\bar{p}, K_{S}^0, \Lambda(\bar{\Lambda})$) and $N_{\rm assoc}^{h^\pm}$ is the number of associated inclusive hadrons. $V_{n\Delta}$ is the two particle harmonic coefficient corresponding to a pair of trigger and associated particles within a given transverse momentum bin ($p_{T}^{\rm trig}, p_{T}^{\rm assoc}$). The details of the procedure to extract the pair correlation function are outlined in the supplementary material. As in the case of experiment, the Fourier anisotropy coefficient $v_{n} (p_{T})$ is defined to be  
\beq
\label{eq_v2e}
v_{n} \{2\} (p_{T}^{\rm trig}) = \frac{V_{n\Delta}(p_{T}^{\rm trig}, p_{T}^{\rm assoc})}{\sqrt{V_{n\Delta}(p_{T}^{\rm assoc},p_{T}^{\rm assoc})}}\,.
\eeq
Our results for $v_{2} (p_{T})$ of identified hadrons are shown in Fig. \ref{fig_v2mass}. A clear mass ordering of $v_{2}$ from light to heavy particles is seen at low momentum. With increasing $p_T$, the mass ordering decreases and is even reversed for some of the species for $p_T > 3$ GeV. Our results for $p+p$ collisions at $\sqrt{s}=7$ TeV are qualitatively similar to those presented by the CMS collaboration at $\sqrt{s}=7$ TeV~\cite{CMS:2015kua} and at $\sqrt{s}=13$ TeV~\cite{Khachatryan:2016txc}. 

While the mass splitting and $N_{\rm ch}$ dependence of $\la p_T \ra$ has been reproduced previously in the initial state PYTHIA color-reconnection scheme~\cite{Ortiz:2013yxa}, that of $v_2(p_T)$ has not.  { It was previously speculated that initial state correlations and fragmentation effects can lead to species dependences of $v_2$~\cite{Torrieri:2013aqa}.  In semi-quantitative studies, another non-hydrodynamic approach, for light-heavy ion collisions~\cite{Romatschke:2015dha}, reproduces $\la p_T\ra$ but finds very small values of $v_2(p_T)$. Likewise, the mass ordering of $v_2(p_T)$ is obtained for light-heavy ion collisions in hadron transport~\cite{Zhou:2015iba} and parton transport~\cite{Li:2016ubw} approaches; they both conclude that mass ordering is not a feature of the scattering but of hadronization, however {their agreement with data requires dominance of final state rescattering}.}

{The  results shown in Figs.~\ref{pid_avgpt} and \ref{fig_v2mass} demonstrate for the first time that, for $p+p$ collisions, the mass ordering pattern of  i) $\la p_T \ra$ (and its $N_{\rm ch}$ dependence) and ii) $v_2(p_T)$ seen in data can both be reproduced in an ab initio initial state framework that does not rely on hydrodynamic flow or final state rescattering.}

The CGC+Lund event generator developed in this paper provides the basis for further phenomenological studies. In addition to more quantitative modeling of data, we will address in the future whether this approach can describe the convergence of the  $m$-particle anisotropic Fourier coefficients $v_n\left\{m\right\}$ seen for large multiplicities, as well as the extension of this framework to describe the systematics of light-heavy ion collisions. 

\textit{Acknowledgements:} We thank Christian Bierlich, Ilkka Helenius, Wei Li, Subhash Singha and Takahito Todoroki for important discussions. This work is supported in part by the U.S. Department of Energy, Office of Science under contract No. DE- SC0012704. RV thanks the Institut f\"{u}r Theoretische Physik, Heidelberg for their kind hospitality and the Excellence Initiative of Heidelberg University for support. This research used resources of the National Energy Research Scientific Computing Center, a DOE Office of Science User Facility supported by the Office of Science of the U.S. Department of Energy under Contract No. DE-AC02-05CH11231. BPS acknowledges a DOE Office of Science Early Career Award. SS gratefully acknowledges a Goldhaber Distinguished Fellowship from Brookhaven Science Associates. 

\begin{figure}[htb]
\includegraphics[width=0.4\textwidth]{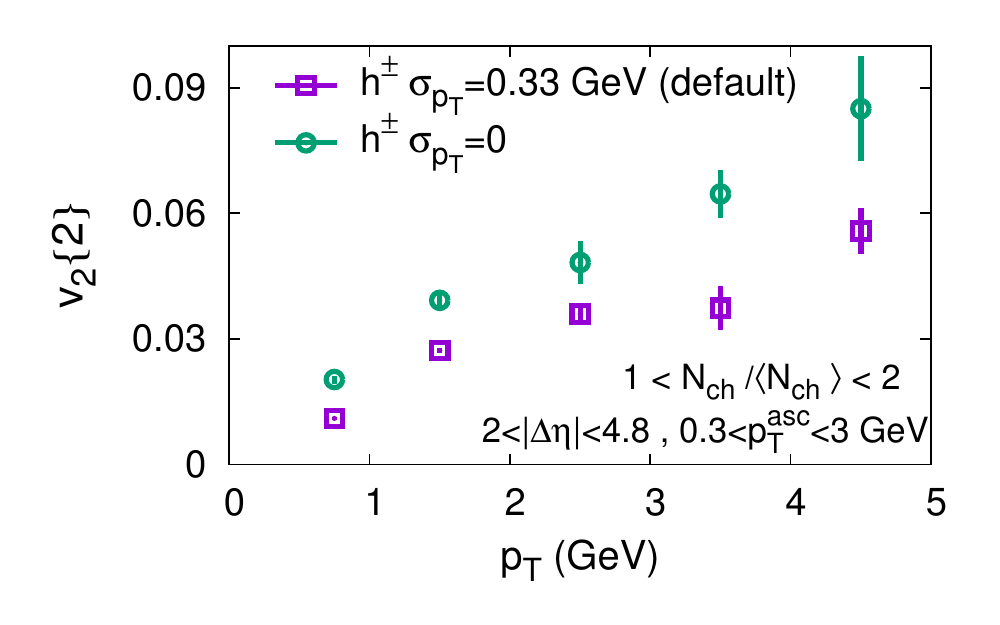}\vspace{-0.2cm}
\caption{Illustration of the effect of varying the transverse momentum smearing fragmentation parameter on $v_2\{2\}(p_T)$.\label{fig_v2sigmapt}}
\end{figure}

\textit{Appendix A:} While no attempt has been made to tune the default parameters of PYTHIA to improve the agreement with experimental data, we will briefly comment on the  sensitivity of our results with respect to the variation of some of the parameters. The most significant effect on two particle correlations comes from changing the transverse momentum smearing parameter $\sigma_{p_T}$ in PYTHIA. 
As an example, the behavior of $v_2\{2\}(p_T)$ for charged hadrons under variations of $\sigma_{p_T}$ is shown in Fig.\ref{fig_v2sigmapt}. Varying $\sigma_{p_T}$ from its default value of  $\sigma_{p_T}=0.33$ GeV to $\sigma_{p_T}=0$ changes $v_2\{2\}(p_T)$ by a factor of $1.5$-$2$ within the range of $p_T$ shown here. This suggests that, with a moderate tuning of parameters, a quantitative comparison of our results with the experimental data can be achieved.

\textit{Appendix B:} The two particle correlation function as a function of the relative difference in pseudo-rapidity $\Delta\eta$, and in azimuthal angle $\Delta\phi$, of a hadron pair is defined in experiments to be 
\beq
\frac{1}{N_{\rm trig}} \frac{dN^{\rm pair}}{d\Delta\eta d\Delta\phi} = B(0,0) \times \frac{S(\Delta\eta, \Delta\phi)}{B(\Delta\eta \Delta\phi)}. 
\eeq
where $S(\Delta\eta, \Delta\phi)$ and $B(\Delta\eta, \Delta\phi)$ are the signal and background pair distributions. These are estimated respectively from real and mixed events. We estimate $S(\Delta\eta, \Delta\phi)$ using a regular event sample after fragmentation. However, unlike the experimental analyses, we do not perform event mixing to estimate $B(\Delta\eta, \Delta\phi)$. Instead, before fragmentation at the gluon level, we make a copy of every event wherein we randomize the azimuthal angle of every sampled gluon within the range $0\!<\!\phi\!<\!2\pi$, while keeping their transverse momentum and rapidity unchanged. Such background events after fragmentation do not contain any of the initial state correlations coming from the IP-Glasma model. However they contain correlations due to resonance decays, although these are suppressed by the large rapidity gap. The 
estimation of the background in this way has the advantage that if any artificial correlation is generated due to the modeling of string topology or the fragmentation process, it will be contained in both $S(\Delta\eta, \Delta\phi)$ and $B(\Delta\eta, \Delta\phi)$ and will be eliminated in the estimation of final observables.

\bibliographystyle{apsrev4-1}
\bibliography{cgcpythia}

\end{document}